\documentclass[aps,prl,twocolumn,superscriptaddress,nofootinbib,nobalancelastpage,showpacs,nobibnotes]{revtex4}
\usepackage{epsfig}

\begin{document}
\title
{Collective polarization exchanges in collisions of photon clouds.}

\author{R. F. Sawyer}\email{sawyer@vulcan.physics.ucsb.edu}
\affiliation{Department of Physics, University of California at
Santa Barbara, Santa Barbara, California 93106}

\begin{abstract}
The one-loop ``vacuum" 
Heisenberg-Euler coupling of four electromagnetic fields can lead to interesting collective effects
in the collision of two photon clouds, on a time scale orders of magnitude faster than one 
estimates from the cross-section and density. We estimate the characteristic time for
macroscopic transformation of positive to negative helicity in clouds that are initially
totally polarized and for depolarization of a polarized beam traversing an unpolarized
cloud.


\end{abstract}

\maketitle

\newcommand{\identity}{\:\mbox{\sf 1} \hspace{-0.37em} \mbox{\sf 1}\,}

Some non-linear aspects of vacuum electrodynamics have been tested in 
experiments on Delbruck scattering \cite{ak1} , i. e. the scattering of a photon off
of the Coulomb field of a nucleus, and in photon splitting \cite{ak2}, also in the nuclear
Coulomb field. 

Essentially, these effects hinge on the one-loop effective Lagrangian density for processes in which four or more electromagnetic fields, of long wavelength compared to the electron Compton wavelength, come
together, as described by the Heisenberg-Euler interaction \cite{he} \cite{js}, the fourth order term of which is,
\begin{equation}
L_{I}= \int d^3 x {2 \alpha ^2 \over 45 m^4}[(\bf E^2-B^2)^2+7(E \cdot B)^2 \rm]\,\,, 
\label{he}
\end{equation}
 where $\alpha$ is the fine structure constant and $m$ is the mass of the electron.
\footnote{We use units $\hbar=c=1$ throughout.}
  
The validity of the effective interaction term (\ref{he}), for long wave-length fields, can hardly be doubted.
Nonetheless, its confirmation in an actual photon-photon scattering experiment would be a milestone
of a kind.  If one puts in the numbers for photon-photon scattering itself,
the cross-section is far too small to be measured with current technology.
Indeed the ``light by light" scattering discussed in the very interesting experiment
reported in ref.\cite{slac} was the reaction $\gamma +\gamma \rightarrow e^+ +e^- $,
and does not test vacuum QED at the one loop level. However, Kotkin and Serbo \cite{ks} have pointed
out that a photon of one plane polarization passing through a cloud of photons that are polarized in a different direction in a frame in which the collisions are head-on, will experience a polarization precession with an angular frequency,
$\Gamma_{ p}=4 \alpha ^2 n_\gamma  \omega \omega_c/(15 m^4)$, 
where $\omega$,$\omega_c$ are the respective frequencies of the impinging photon and the cloud and $n_\gamma$
is the number density of cloud photons. This rate is to
be contrasted to the ordinary scattering rate of the impinging photon, as derived from the cross-section
\cite{he}, $\Gamma_s=
.014 \times \alpha^4 m^{-8} n_\gamma \, \omega^3 \omega_c^3$.
In all situations in which $\omega \, \omega_c<<m^2$, $\Gamma_p$
is many orders of magnitude greater than $\Gamma_s$. 

This polarization precession, from an effective anisotropic index of refraction, originates in the coherent interaction through forward
scattering of a single
beam photon with a large number of cloud photons. In this note we develop
the theory of another collective interaction, now between two clouds of photons, also depending on coherent
forward scattering. This interaction can lead to helicity changes 
when photons of both clouds initially all have the same helicity, and to
depolarization of one cloud when the other cloud is initially unpolarized. The rate will now turn out to be of order  $\Gamma_{\rm pol}$  divided by a slowing factor $\log (N)$ where N is the number of photons in a region of interaction of linear dimension
$1/\Gamma_p$.

To rederive the  Kotkin and Serbo result, and to lay the groundwork for the extension, we consider the complete set of momentum states $\{q_i \}$ that are occupied in the initial state in either cloud (whether singly or multiply occupied).  We take $L_I$ of (\ref{he}) and truncate it by keeping the parts of the fields that contain only creation and annihilation operators for this
set of momenta. The momentum-conserving processes described by this interaction
are just the forward scattering of beam photons from cloud photons since co-moving cloud
particles (or beam particles) do not scatter from each other in the interaction, (\ref{he}). The result, for the effective
``forward" Hamiltonian of the system, after substitution of the canonical
expressions for the electromagnetic fields in terms of creation and annihilation operators into $ -L_I$ of (\ref{he}) and performing the space integral over a quantization volume, $V$, is,
\begin{eqnarray}
H_{\rm for}=G V^{-1}\sum_{j,m} \omega_j \omega_m [\zeta^{(1)}_j \tau^{(1)}_m + \zeta^{(3)}_j \tau^{(3)}_m
\nonumber\\
 -(11/3) \identity_j^{(a)}
 \identity_m^{(b)}] \,\,,
\label{forward}
\end{eqnarray}
where $G=2 \alpha^2/15\ m^4 $, and where the indices $j$ and $m$ extend over the momentum states defined above.

In (\ref{forward}) the products of photon annihilation and creation operators for the beam modes, $a^x_j$, $a^y_j$, 
 and for the cloud modes, $b^x_j$, $b^y_j$ (where $x$ and $y$ indicate the
polarization state and $j$ enumerates the set of momenta) have been reexpressed in terms of the operators, 
\begin{eqnarray}
\identity_j^{(a)} =(a^{(x)}_j)^\dagger a^{(x)}_j + (a^{(y)}_j)^\dagger a^{(y)}_j\,\,,
\nonumber\\
  \tau^{(1)}_j=(a^x_j)^\dagger a^y_j + (a^y_j)^\dagger a^x_j \,\,,
\nonumber\\
\tau^{(3)}_j=
(a^{(x)}_j)^\dagger a^{(x)}_j - (a^{(y)}_j)^\dagger a^{(y)}_j \,\,.
\label{operators}
\end{eqnarray} 
with the parallel set of definitions for the cloud operators, taking $a\rightarrow b$, $\vec \tau \rightarrow \vec \zeta$.
The operators $\tau^{(1),(3)}/2$, supplemented by an operator $\tau^{(2)}/2$, which will not explicitly enter below,
obey angular momentum commutation rules, as do the operators $\vec \zeta /2$.
Since $H_{\rm for}$ only connects states of identical \underline{unperturbed} energies, we did not include
a contribution from an $H_0$ in (\ref{forward}). \footnote{We have made the transition from interaction Lagrangian to
the interaction Hamiltonian in (\ref{forward}) without regard to the fact that the original $L_I$ contains time derivatives
of the original canonical coordinates, $\vec A$. This is consistent if we take $\omega_j$ and $\omega_m$ as simple
parameters in $(\ref{forward})$ and do not translate them back into time derivatives when we go to a Heisenberg
picture.}  
To follow the polarization of a single beam photon of energy $\omega$ passing through the cloud we can then write the
Heisenberg equations for $\vec \tau (t)$ coming from (\ref{forward}) as,
\begin{equation}
{d \over dt}\vec \tau (t)=-2 G \omega V^{-1} \,\vec \tau(t) \times (\vec Z (t)-\vec v Z^{(2)})\,.
\label{onephoton}
\end{equation}
where we have defined $\vec Z=\sum_m \omega_m \vec \zeta_m$ , and introduced a vector $\vec v$, defined as a unit vector in the $2$ direction in the internal space \footnote{One should not confuse the three dimensional space that we have
introduced using a spinorial representation for the polarization vectors with the three dimensional configuration space; we label
vector components in the former with (1,2,3) and in the latter with (x,y,z)}.
For the case of an isolated beam photon interacting collectively with the cloud photons it
is fairly clear that we can replace the cloud operators, $Z^{(1),(3)}$ by their expectation values,
since the back reaction from beam interactions affects the cloud almost not at all. If the cloud
polarization is at an angle $\theta$ to the $\hat x$ axis and the cloud energies are reasonably narrowly
clustered around an energy $\omega_c$ we have $\langle Z^{(3)}\rangle / V =\omega_c n_{\gamma} \cos (2 \theta)$, 
$\langle Z^{(1)} \rangle /V=\omega_c n_{\gamma} \sin (2 \theta)$. Since the eqs.(\ref{onephoton}) are now linear in
the operators for the beam particle, they hold for expectation values. Taking the initial condition  
$\langle \tau^{(3)}(0) \rangle =1$, $\langle \tau^{(1),(2)}(0) \rangle =0$, for an initial beam polarization
in the $\hat x$ direction, and solving (\ref{onephoton}), we obtain the
the $x,x$ component of the polarization density matrix, $P_x=1/2+\langle \tau^{(3)}(t)\rangle/2 $,
\begin{equation}
P_x=1-{1 \over 2}\sin^2 (2 \theta)[1- \cos (\Gamma_p t)]\, ,
\label{osc}
\end{equation}
where $\Gamma_p=2 G\omega \omega_c n_\gamma$.

Eq.\ref{osc} recaptures the effects noted by Kotkin and Serbo \cite{ks}, and we refer the reader to their articles
for more discussion as to the possibilities of observations. To make one comparison to laboratory
parameters, we define oscillation length 
as $\lambda=(\Gamma_p)^{-1}$ and express in ordinary units, 
\begin{equation}
\lambda=1.5 \times 10^{-9}\Bigr ( {E_{\rm crit} \over  E} \Bigr )^2 \Bigr ({ 1 {\rm MeV}\over \hbar  \omega}\Bigr )\,\, {\rm cm}\,\, ,
\end{equation} 
where $E_{\rm crit}=m^2c^3/e \hbar$ and $E$ is the rms electric field of the cloud.
In the $\omega_1=2.35 {\rm eV}$ laser used in the experiment
reported in ref.\cite{slac} the field strength was $ E / E_{\rm crit}\approx  1.5 \times 10^{-6} $. In this case taking $\hbar \omega = 100 {\rm MeV}$ leads to an oscillation length of $\approx 3 \,\,{\rm cm}$ (The pulse length for this laser is a fraction of a millimeter; the free path for ordinary photon scattering from the cloud under these conditions is of the
order of $10^9$ cm.)

We further note that this photon-cloud interaction produces no effect, on the short time scale, if the inital polarizations
are perpendicular, and we note that if the cloud is unpolarized then there is no depolarization of the beam. 
Turning to the case of two colliding clouds, for which neither of these conclusions will hold, 
we assume for simplicity that photon densities in the two colliding groups 
are equal. Now we need to take the variables $\vec \zeta$ on the R.H.S. of (\ref{forward}) as well as the variables
$\vec \tau$ to be dynamic variables,
rather than taking their expectation values in the initial state.

This calculation is simplest in a helicity basis, however.
The forward interaction, $H_{\rm for}$, gives a matrix
element for the transition in which a state of a positive helicity photon from one
bath and a positive helicity photon from the other bath makes a transition to a state with two photons of negative helicities. We can easily express $H_{\rm for}$
of (\ref{forward}) now in terms of operators $\vec \xi ,\,  \vec \eta$ which act in the two dimensional helicity spaces of the respective clouds, designated respectively as the ``up" cloud and
the ``down" cloud. The components $\xi_i^{(3)}$ and $\eta_i^{(3)}$ measure the spins in the $\pm \hat z$ direction for the photons in the respective clouds, thus the negative of the helicity in the case of the down-moving photon. We choose both clouds to be essentially monoenergetic, with energies $\omega$ and $\omega_c$ for the
respective up-moving and down-moving clouds; then we can express the ``forward" Hamiltonian in terms of the
collective coordinates,
$\xi^{(\pm)}=\sum_i \xi_i^{(\pm)}$, $\eta^{(\pm)}=\sum_i \eta_i^{(\pm)}$, where $\xi^{(+)}=(\xi^{(1)}+i\xi^{(2)})/2$ etc.

By direct transformation of  (\ref{forward}) we obtain,
\begin{eqnarray}
H_{ \rm for}= G \omega \omega_c V^{-1} [ 2 \xi ^{(+)} \eta ^{(-)}+2\xi ^{(-)}\eta ^{(+)} 
\nonumber\\  -(11/3) \identity ^{(a)} \identity ^{(b)} ] \, .
\label{laser}
\end{eqnarray}

Now we pose the question of what happens
beginning with an initial state in which all N up-moving photons have spin $+1$ and
all N down-moving photons have spin $-1$ in the $\hat z$ direction.  We can proceed,
as in the earlier case, by writing the equations of motion,
\begin{eqnarray}
{d \over dt} \xi^{(+)}(t) =-2i G \omega \omega_c V^{-1}\xi^{(3)}(t) \eta^{(+)}(t) \, ,
\nonumber\\
{d \over dt} \xi^{(3)}(t) =-2iG \omega \omega_c V^{-1}[\xi^{(+)}(t) \eta^{(-)}(t)-\xi^{(-)}(t) \eta^{(+)}(t)],
\nonumber\\
{d \over dt} \xi^{(-)}(t) =2i G \omega \omega_c V^{-1}\xi^{(3)}(t) \eta^{(-)}(t)~\, ,~~~~~~
\label{eom2}
\end{eqnarray}
plus the three equations in which $\vec \tau$ and $\vec \zeta$ in (\ref{eom2}) are interchanged. 
In the calculation leading to (\ref{osc}) we proceeded to a soluble problem by taking 
a factorized ansatz that is equivalent, in our present problem, to the replacement,
\begin{equation}
\langle \xi^{(3)}(t) \eta^{(+)}(t)\rangle =\langle\xi^{(3)}(t) \rangle \langle  \eta^{(+)}(t) \rangle \, \, .
\end{equation}
But for the initial state that we are now considering all of the mixing operators
with  $\pm$ superscript have expectation value zero, and it is clear that there would be no evolution in time at
all, were the factorization ansatz valid. We proceed instead to a calculation equivalent to solving
the full coupled operator equations.

The total $\hat z$ component angular momentum in the
new internal space in which helicity is the basis, measured by $ (\xi^{(3)}+ \eta^{(3)})/2$ , is  conserved. 
Thinking of the system as an assemblage of spins with an upper tier of $N$ spins all initially pointed
up and a lower tier all initially pointed down, we enumerate the states that are connected to the initial state
(and to each other) by the Hamiltonian of (\ref{laser}). Any number of the $N$ spins in the upper tier, all initially up, may be flipped, leading to $ N+1$ possibilities for the magnetic quantum number of the this tier. The operators $\vec \xi \cdot \vec \xi /4$
and  $\vec \eta \cdot \vec  \eta /4$ are separately conserved, each with eigenvalue $(N/2+1)N/2$. Therefore for each value of 
$ (\xi^{(3)}/2)$ in our set, there is a single upper tier configuration that enters, and a single lower tier
configuration as well. We index the states by the number of flips plus one, $i$, where $i$ takes on the values $1,2...N+1$. We express the operator products that occur in the Hamiltonian in this basis,
\begin{eqnarray}
\langle i | \xi_-\eta_+| i-1\rangle = (N-i+1)(i) ;\,i=1,..N+1\, ,
\nonumber\\
\langle i+1 |\xi_+\eta_-| i\rangle=(N-i+2)(i-1);\,i=1,..N+1 ,
\label{bmats}
\end{eqnarray}
which come directly from the standard angular momentum matrices. We solve numerically for 
a $n+1$ component wave function $\Psi(t)$, using the Hamiltonian (\ref{laser}) with the substitution (\ref{bmats})
and the initial condition 
$\Psi_i(0)=\delta_{i,1}$, and then calculate the measure of average helicity of the upper tier,
\begin{equation}
R(t)=N^{-1}\sum_{i=i}^{N} \langle  \xi_3^{(i)}\rangle =\sum_{i=1}^{N+1} |\Psi_i(t)|^2 (N-2i+2) N^{-1}.
\label{ans}
\end{equation}
We perform these calculations for a series of values of N, and show the results as a function of scaled time, $s=\Gamma_p \,t=2GN\omega_c \omega t/V$. Fig.1 displays results for values of N ranging from $8$ to $512$, equally spaced in $\log (N)$. 

\begin {figure}[ht]
    \begin{center}
        \epsfxsize 2.75in
        \begin{tabular}{rc}
            \vbox{\hbox{
$\displaystyle{ \, { } }$
               \hskip -0.1in \null} 
} &
            \epsfbox{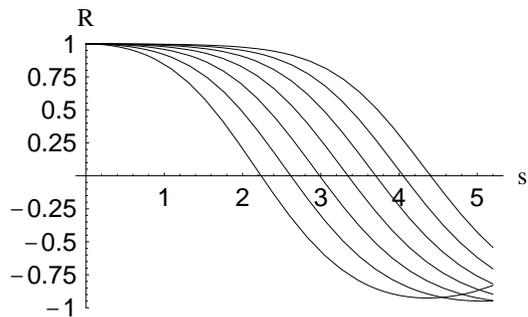} \\
            &
            \hbox{} \\
        \end{tabular}
    \end{center}
\label{fig.1}
\protect\caption
    {%
The function $R(s/ \Gamma_p)$ of  (\ref{ans}), the mean helicity of the up-moving
cloud, for values of $N=8,16,32,64,128,256,512$ as determined from
solutions of (\ref{eom2}), plotted against the dimensionless scaled time, $s$
The curves for higher values of $N$ lie progressively farther to
the right. Equal spacings of the curves indicates a transition time increasing as
$\log(N)$.
 }
\end {figure}

The data shown in the
figure clearly suggests a characteristic time of order $\Gamma_p^{-1}\log (N)$ for a complete turnover of the spins.
We can gain a heuristic understanding of these results. Instead of the set of operators $\vec \xi, \, \vec \eta$ we introduce
the bilinear forms,
\begin{eqnarray}
x=i\xi^{(+)} \eta ^{(-)}~;~ u=i\xi^{(-)} \eta^{(+)} ~;~  y=\eta^{(+)} \eta^{(-)},
\nonumber\\ z=\xi^{(-)} \xi^{(+)} ~~;~~ w=\xi^{(3)}~~~
\label{fiveops}
\end{eqnarray}
Writing the Heisenberg equations of motion for these operators by taking commutators
with $H_{\rm for}$ in the form (\ref{laser}) and making the further substitution $\eta ^{(3)}=-\xi ^{(3)}$
we obtain the closed set,
\begin{eqnarray}
N\dot x=w(z+y)-w^2 ~~;~~\dot u=-\dot x ~~;~~N \dot y=w x-u w,
\nonumber\\
N \dot z =xw-wu~~;~~N \dot w=-x+u ~.~~~~~~~~~
\end{eqnarray}
where the derivatives are with respect to the scaled time $s$.
Treating these equations as c-number equations \footnote{This is exactly the arbitrary procedure that we disparaged above for the case of the
equations for the operators $\vec \xi$ and $\vec \eta$. One difference, and perhaps the key to the agreement
with results of the complete solutions, is that the operators (\ref{fiveops}) keep us within
the $N+1$ dimensional subspace of states defined above, while the operators $\vec \xi$ and $\vec \eta$ do not.}
 with the initial conditions $x=y=z=u=0$ and $w=N$
allows us to write a single equation for $\bar w \equiv w /N$,
\begin{equation} 
{d^2 \over ds^2} \bar w =2 \bar w( \bar w^2 -1) +{2 \bar w^2 \over N}\, .
\label{soliton}
\end{equation}
The initial condition is now $\bar w(0)=1$, $\bar w ' (0)=0$. In fig.2 we plot solutions of (\ref{soliton}) and compare with the numerical solutions to the complete equations.
\begin {figure}[ht]
    \begin{center}
        \epsfxsize 2.75in
        \begin{tabular}{rc}
            \vbox{\hbox{
$\displaystyle{ \, { } }$
               \hskip -0.1in \null} 
} &
            \epsfbox{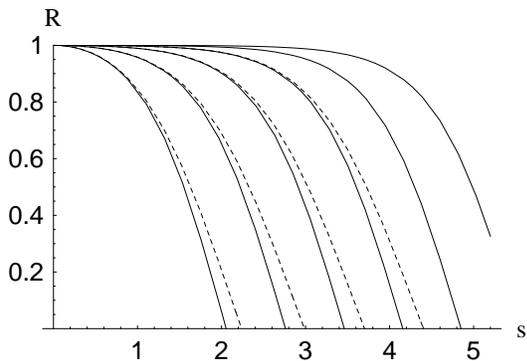} \\
            &
            \hbox{} \\
        \end{tabular}
    \end{center}
\label{fig. 2}
\protect\caption
    {%
The function $R(s/\Gamma_p)$ as determined from the solution for the heuristic
equation (\ref{soliton}) for values $N=8,32,128,512,2048,8192$ (solid
lines). The dashed lines show the solutions of the complete equations of
motion (\ref{ans}), as plotted in fig.1 for the first four values of $N$. 
 }
\end{figure}
The fit is good for values of $R>.6 \,$. 
We  also see that the in the case of solutions to (\ref{soliton}) the equal spacing continues to values of $N \approx 10^4$, leaving little
doubt of the logarithmic dependence. It is possible to understand this limit analytically from (\ref{soliton}),
capitalizing on the fact that when $N\rightarrow \infty $, the solution is the familiar
kink solution in a $\lambda \phi ^4$ theory in one dimension, $\bar w={\rm tanh}[(t-t_0)/2)]$, then 
showing that for large $N$,
in the time region in question, the $ \bar w^2/ N$ term in (\ref{soliton}) can be dropped in favor
of changing the initial value of $w$ to $1-2/N$,  this in turn determining $t_0=\log(2N)$. 

To summarize briefly: in many-body systems in which every partice of set A interacts with every particle of set
B, evolution times for macroscopic properties may be much faster than one would have predicted based
on cross-sections, even in the absence of initial phase relations among the components that one might
have anticipated were necessary for such behavior. In the photon-photon system the effect is an extension
of the known index-of-refraction effects of photon polarization treated in refs. \cite{ks}.  In the detailed example treated, there is total oscillation back and forth between
all positive helicities and all negative helicities in both clouds. 

The case in which one cloud with 100\% polarization in helicity collides with an unpolarized cloud is
somewhat more complex. Here
we predict partial depolarization of the polarized cloud. From (\ref{laser}) we see that photons in the
target cloud with the opposite helicity to those of the beam cloud are effectively sterile. Therefore
we can discuss the polarization changes of the beam cloud in a manner similar to that of the calculation 
given above. There remains an order $\Gamma_p /\log (N)$ rate of depolarization after averaging
over the configurations of polarization of the individual photons in the target cloud.
This is in contrast to case of a single photon interacting with the cloud discussed at the beginning of this paper,
where depolarization takes place on the time scale $1/ \Gamma_s$.

Our calculation was for an idealized system of plane wave modes in a box, with (implicit) periodic boundary conditions. Does it apply to realizable systems in which the two clouds are in contact  for a time of order (box size/c)?
It is clearly required that the characteristic time for transformation be shorter than this contact time, a criterion that
is easily checked in any given situation. It is harder to answer the question, ``Can the laboratory 
photons in the two beams really 
sustain a coherent interaction over the whole of the macroscopic region (of order of a cm., in the numerical example
mentioned above, but now multiplied by a logarithm of the order of 100) for our process to unfold?" We do
not know how to address the exact quantification of this question, although we believe that the answer is ``yes"
for the case of the beams from lasers and from synchrotrons. Another question is that of the role of all of the modes
that we have left out in using the truncation that produced the ``forward" Hamiltonian (\ref{forward}). We
anticipate that over the time-scale $1/\Gamma_p$ these modes create junk that does not add up to anything macroscopically,
due to phase oscillations, as indeed they must in our preliminary beam-cloud calculation. In any case we believe that the
``speed-up" through the many body interactions that we have described here is interesting enough to
warrant serious attention to some of the harder questions that arise.

Finally we note the close similarity between the issues discussed in this note and in refs. \cite{bell},
\cite{fl}, which discussed the possibility of speeded-up flavor transformations of colliding
neutrino clouds. Although the equations are quite similar, a critical difference is a term proportional
to $\xi^{(3)} \eta^{(3)}$ on the RHS of the analogue of (\ref{laser}) in the neutrino case.
This term destroys the speed-up process in the simple
model, with just two tiers of states, in which all the couplings between the upper tier and 
lower tier neutrino states (in our N-spin terminology) are equal to each other. In more
realistic (and complex) situations it is possible that there would be speeded evolution, however.

\end{document}